\documentclass[epj]{webofc}
\usepackage[varg]{txfonts}   
\wocname{EPJ Web of Conferences}
\woctitle{INPC 2013}
\begin{document}
\title{Heavy flavour in nucleus-nucleus collisions at RHIC and LHC: a Langevin approach}

\author{A. Beraudo\inst{1}\fnsep\thanks{\email{Andrea.Beraudo@cern.ch}} \and
        A. De Pace\inst{2} \and
        M. Monteno \inst{2} \and
        F. Prino \inst{2} \and
        W.M. Alberico \inst{2,3} \and
        A. Molinari \inst{2,3} \and
        M. Nardi \inst{2}
}

\institute{Physics Department, Theory Unit, CERN, CH-1211 Gen\`eve 23, Switzerland
\and
Istituto Nazionale di Fisica Nucleare, Sezione di Torino, via P.Giuria 1, I-10125 Torino, Italy             
\and
Dipartimento di Fisica dell'Universit\`a di Torino, via P.Giuria 1, I-10125 Torino, Italy          
          }

\abstract{%
  A snapshot of the results for heavy-flavour observables in heavy-ion ($AA$) collisions at RHIC and LHC obtained with our transport calculations is displayed. The initial charm and beauty production is simulated through pQCD tools (POWHEG+PYTHIA) and is validated through the comparison with data from $pp$ collisions. The propagation of $c$ and $b$ quarks in the medium formed in heavy-ion collisions is studied through a transport setup based on the relativistic Langevin equation. With respect to past works we perform a more systematic study, providing results with different choices of transport coefficients, either from weak-coupling calculations or from lattice-QCD simulations. Our findings are compared to a rich set of experimental data ($D$-mesons, non-photonic electrons, non-prompt $J/\psi$'s) which have meanwhile become accessible.
}
\maketitle

\section{Introduction}
\label{sec:intro}
The heavy-ion program of all the detectors at RHIC and LHC nowadays include different heavy-flavour measurements which, all together, provide a rich experimental information challenging the various theoretical models to consistently reproduce such a wide set of data.

Heavy quarks, produced in the initial stage in hard pQCD processes (under good theoretical control), after comparing the findings in $pp$ and $AA$ collisions allow one to perform a tomography of the Quark Gluon Plasma (QGP) produced in heavy-ion experiments. At the same time the detailed information on the medium provided independently by hydrodynamic studies of soft observables can be exploited to put tight constraints on the interaction between the heavy quarks and the plasma, which must lead to the observed amount of quenching and azimuthal asymmetry of the final spectra.  

From the theory side, the study of heavy quarks in $AA$ collision requires to develop a multi-step setup: the simulation of their initial production, possibly including initial-state nuclear effects; a realistic modeling of the medium; the description of their dynamics in the QGP until hadronization, which -- taking place in the presence of a medium -- represents a further source of systematic uncertainty; finally, the simulation of the decays into the final observables (e.g. $D\to X\nu e$, $B\to X J/\psi$...). The fact that each model addresses the above issues in a different (sometimes schematic) way makes a comparison of the different calculations difficult and the question of constraining the interaction between the heavy quarks and the rest of the plasma hard to answer.

Here we present some selected outcomes of the transport setup developed by our team in the past~\cite{EPJ1} and recently updated~\cite{EPJ2} to include a more realistic simulation of the initial heavy-quark production, a wider systematic study of the heavy-flavour transport coefficients and a comparison with the new experimental observables which have become meanwhile available.

For independent approaches we refer the reader to Refs.~\cite{rapp,aka,aic2,BAMPS}.
\section{The setup}
\label{sec:setup}
Our setup for heavy-flavour studies, described in detail in~\cite{EPJ1,EPJ2}, includes several steps. Heavy-quark ($c$ and $b$) production is simulated through the POWHEG-BOX package~\cite{POWBOX}, which provides a user-friendly frame to interface the hard-event, evaluated at NLO accuracy, to a shower stage simulated through PYTHIA. In the $AA$ case nuclear effects (nPDFs and Cronin broadening) are also included. Heavy quarks are then made propagate in the medium produced in the collisions, information on the latter being provided by the output of hydrodynamic codes~\cite{rom1}. Their stochastic motion is described through the relativistic Langevin equation (hence the acronym POWLANG for our setup)
\begin{equation}
  \frac{\Delta{\vec p}}{\Delta t}=-\eta_D(p)\vec{p}+\vec{\xi}(t)\quad{\rm with}\quad\langle\xi^i(t)\xi^j(t')\rangle=b^{ij}(\vec{p})\delta_{tt'}/\Delta t,
\label{eq:lange_r_d}
\end{equation}
in which the evolution of the heavy quarks is driven by a deterministic friction force and by a stochastic noise term, fixed by its two-point correlator. The latter is local in time (collisions at different time-steps taken as uncorrelated ) and, after decomposing the tensor $b^{ij}(\vec{p})\!\equiv\! \kappa_L(p)\hat{p}^i\hat{p}^j\!+\!\kappa_T(p)(\delta^{ij}\!-\!\hat{p}^i\hat{p}^j)$, turns out to depend on two transport coefficients, $\kappa_{T/L}$, which have to be calculated from the underlying theory and, from the physics point of view, represent the average squared-momentum exchange per unit time with the medium. In the following we will compare results obtained in weak-coupling~\cite{EPJ2} and in lattice-QCD~\cite{hflat1} calculations. Finally, around a critical energy-density, heavy quarks are made hadronize via standard in-vacuum fragmentation, as in elementary collisions, and the relevant decay-chains (into electrons from semi-leptonic decays or non-prompt $J/\psi$'s from B) are also simulated.
\section{Results}
\label{sec:results}
\begin{figure}
\begin{center}
\includegraphics[clip,width=0.48\textwidth]{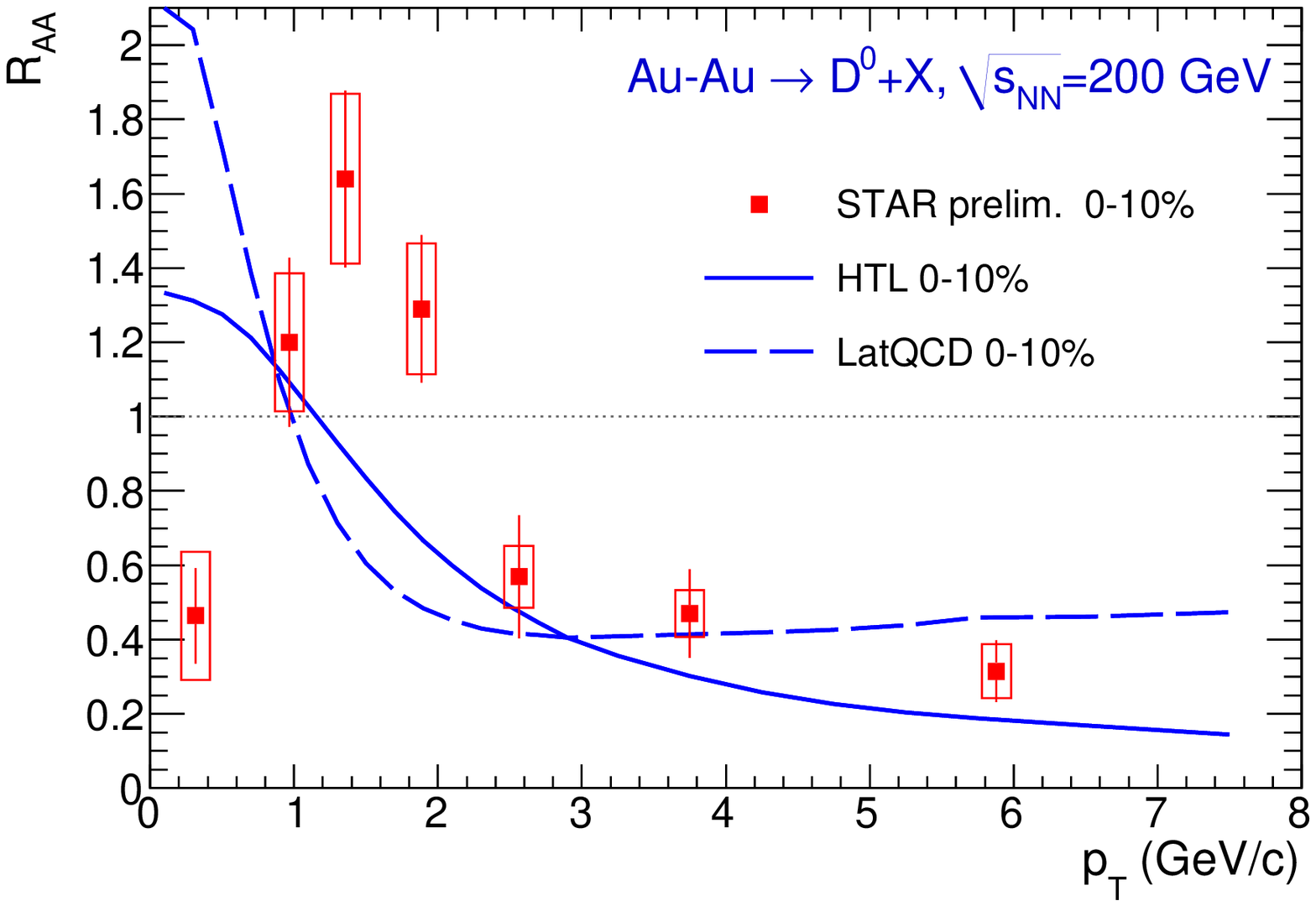}
\includegraphics[clip,width=0.48\textwidth]{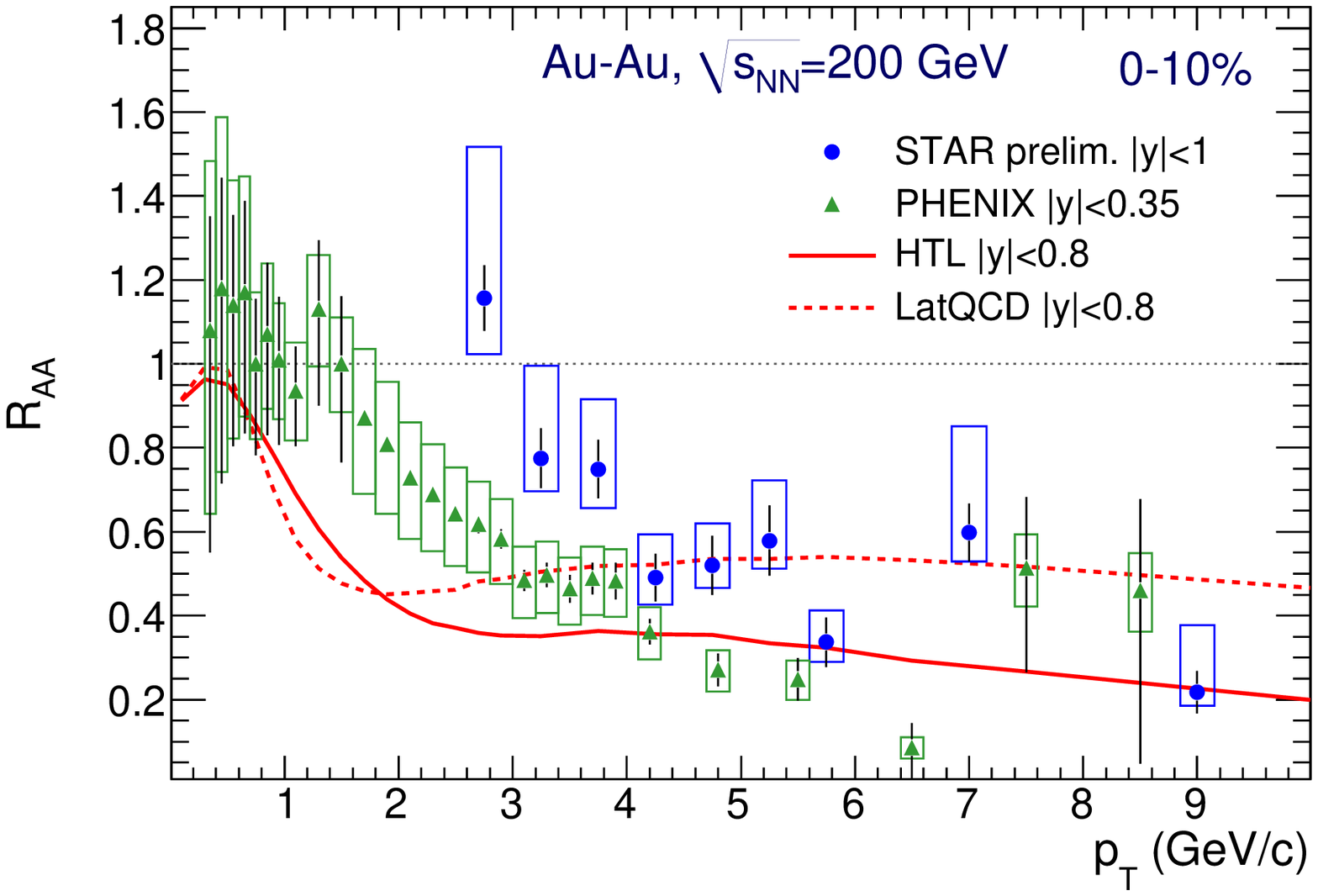}
\caption{Results (with HTL and lattice-QCD transport coefficients) for the
  $R_{AA}$ of $D^0$ mesons (left panel) and heavy-flavour electrons (right panel) in central ($0\!-\!10\%$) Au-Au collisions at $\sqrt{s_{NN}}\!=\!200$ GeV. Our findings are compared to preliminary data from the STAR~\cite{STAR_D,STARhfe} and PHENIX~\cite{PHE} collaborations.}
\label{fig:RAA_STAR}
\end{center} 
\end{figure}
POWLANG results are compared in Fig.~\ref{fig:RAA_STAR} to data for the $R_{AA}$ of $D^0$ mesons and non-photonic electrons (from $c$ and $b$ decays) in Au-Au collisions from the STAR and PHENIX collaborations. One can observe an overall agreement, although the peak in the $D^0$ $R_{AA}$, possibly coming from coalescence (not included in our setup), requires a deeper investigation.

\begin{figure*}
\begin{center}
\includegraphics[clip,width=0.48\textwidth]{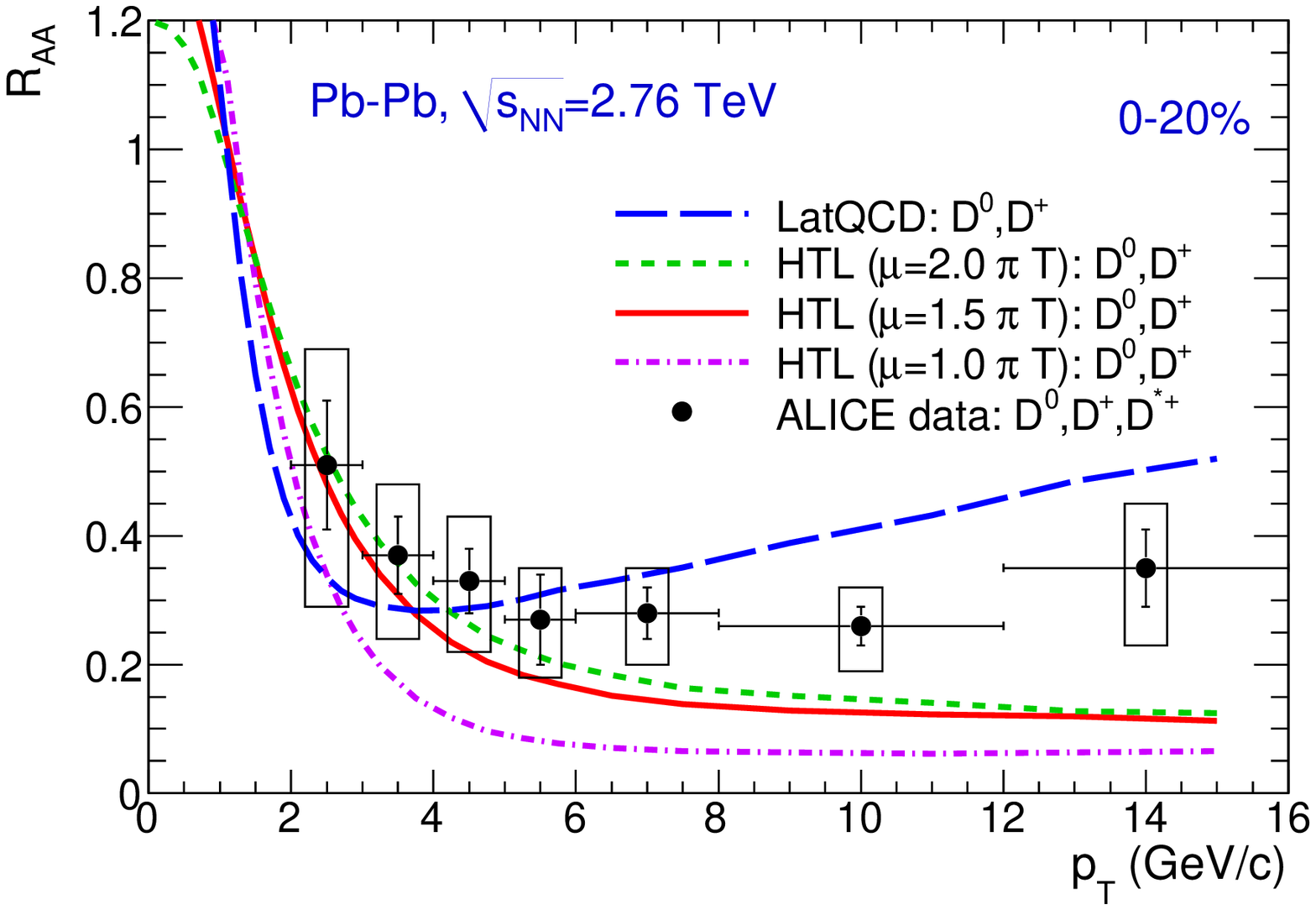}
\includegraphics[clip,width=0.48\textwidth]{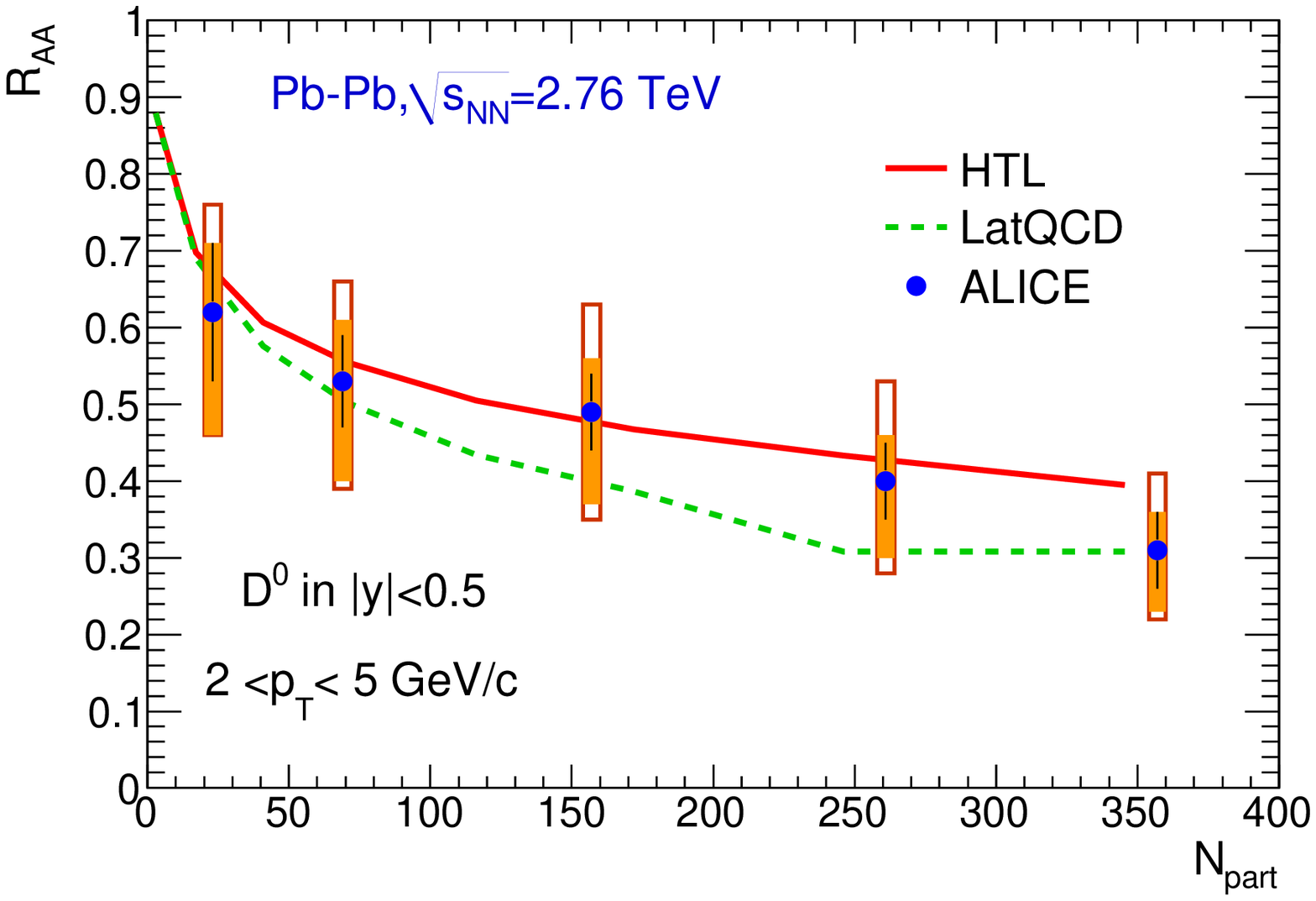}
\caption{$D$-meson nuclear modification factor versus $p_T$ (left panel) and centrality (right panel) in Pb-Pb collisions at $\sqrt{s_{NN}}\!=\!2.76$ TeV. Results with HTL (with different scales of the coupling in the left panel) and lattice-QCD transport coefficients are displayed and compared to ALICE data~\cite{ALI_D}.}
\label{fig:RAA_D}
\end{center}
\end{figure*}
\begin{figure*}
\begin{center}
\includegraphics[clip,width=0.48\textwidth]{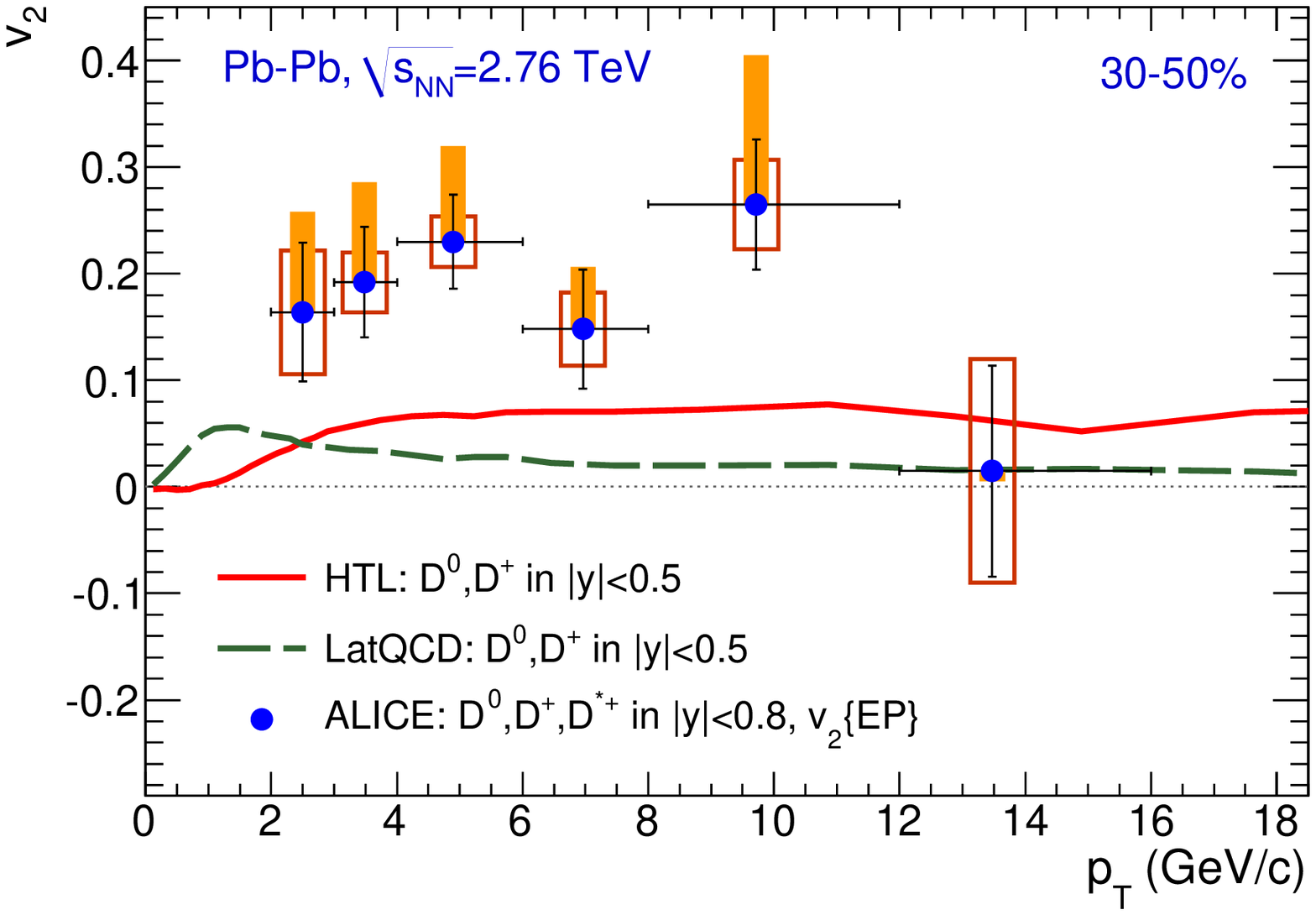}
\includegraphics[clip,width=0.48\textwidth]{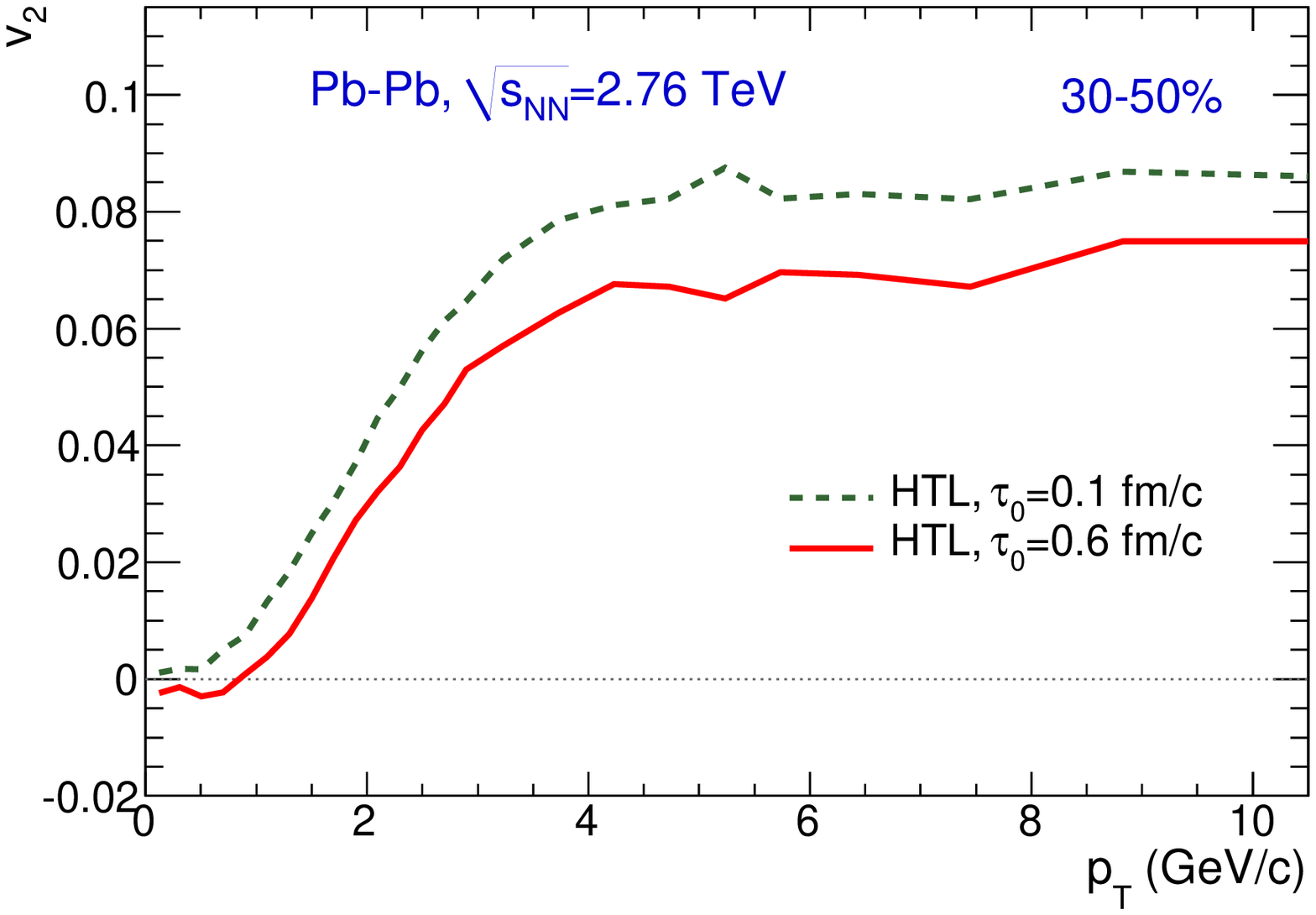}
\caption{Left panel: elliptic flow of $D$ meson in semi-peripheral
  ($30\!-\!50\%$) Pb-Pb collisions at the LHC compared to
  ALICE data~\cite{ALIv2D}. Results with HTL and
  lattice-QCD transport coefficients are displayed. 
  Right panel: dependence of the $v_2$ on
  the thermalization time $\tau_0$ of the background medium.}
\label{fig:v2_D}
\end{center}
\end{figure*}
Figs.~\ref{fig:RAA_D} and~\ref{fig:v2_D} refer to the charm measurements by ALICE in Pb-Pb collisions at the LHC. Our model manages to reproduce reasonably well the quenching of the spectra of $D$-mesons (for not too large $p_T$). On the other hand, with the same transport coefficients, we underestimate the observed elliptic flow, even with the extreme choice for the initial time of the medium evolution $\tau_0\!=\!0.1$ fm/c. Hopefully the agreement with the data might improve after inclusion of hadronization via coalescence (currently in progress), the light quark providing also its own contribution to the meson $v_2$. 

\begin{figure*}
\begin{center}
\includegraphics[clip,width=0.48\textwidth]{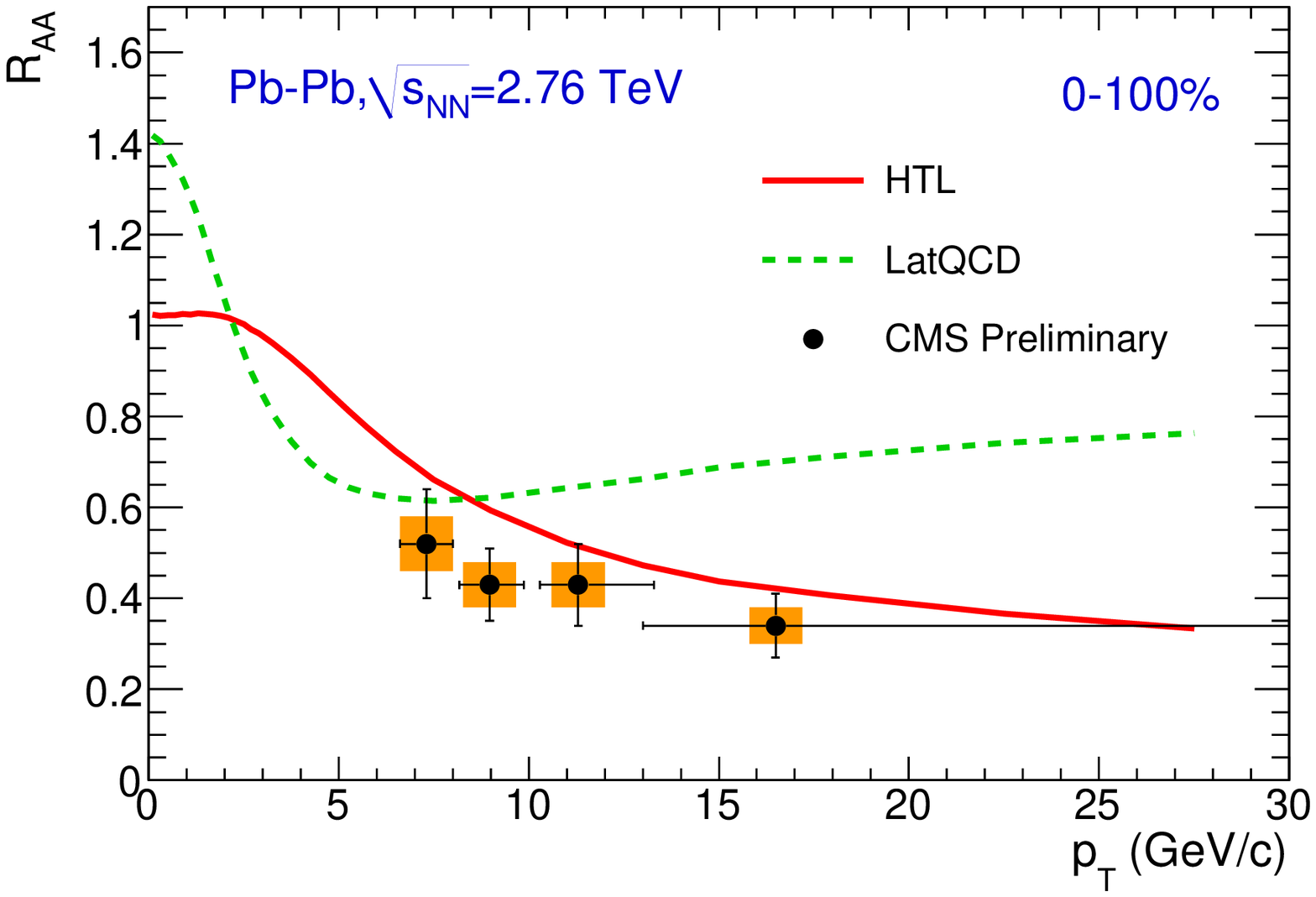}
\includegraphics[clip,width=0.48\textwidth]{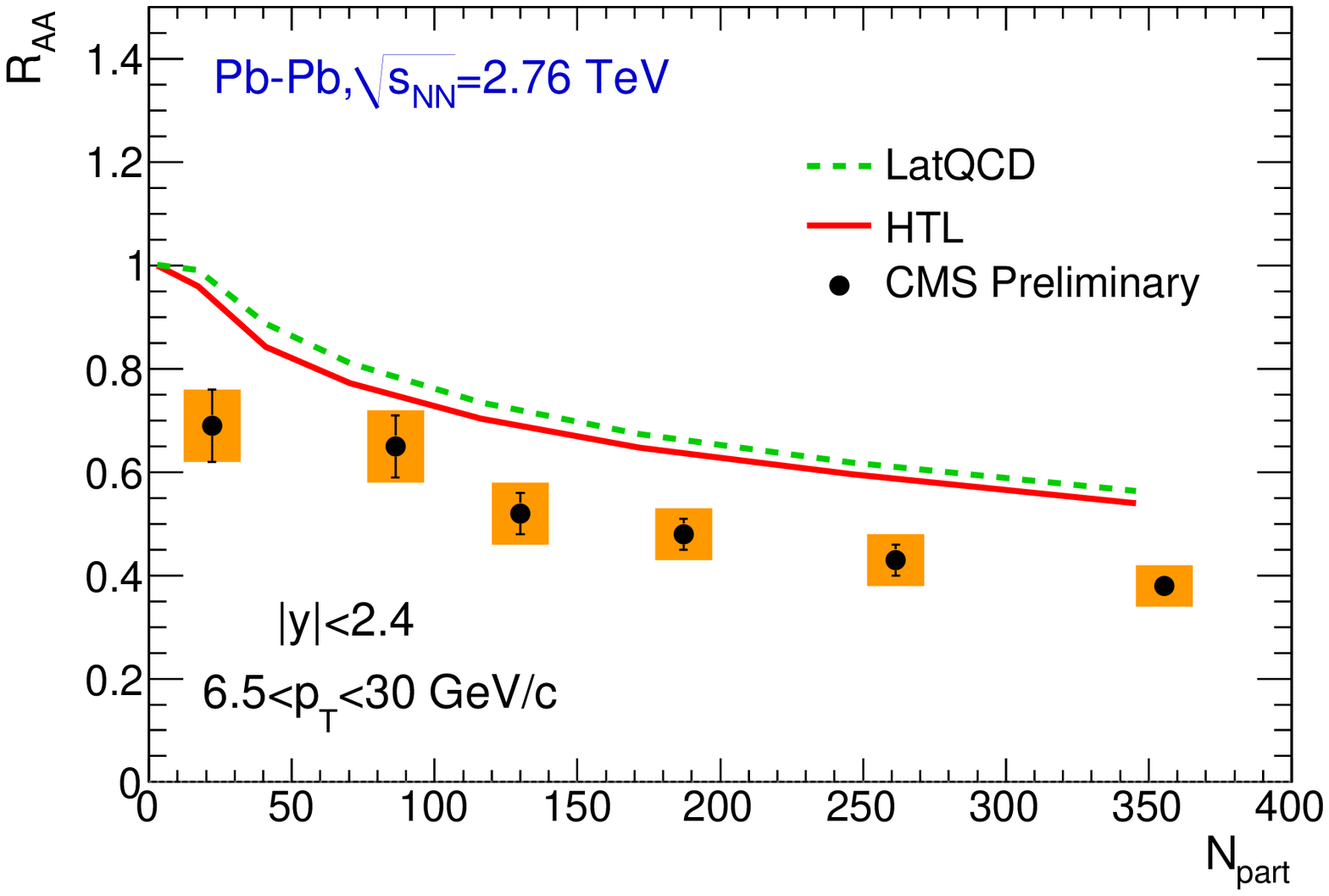}
\caption{Left panel: $R_{AA}$ as a function of $p_T$ of non-prompt $J/\psi$'s
  (from B decays) in minimum-bias Pb-Pb collisions at the LHC. 
  Results of our setup (with HTL and lattice-QCD transport coefficients) are
  compared to preliminary CMS data \cite{CMSJpsi}. Right panel: the centrality
  dependence of the $R_{AA}$ of non-prompt $J/\psi$'s.}
\label{fig:RAA_Jpsi}
\end{center}
\end{figure*}
Finally, in Fig.~\ref{fig:RAA_Jpsi} we display our findings for the quenching of the non-prompt $J/\psi$'s from $B$ decays, currently representing the only way to access information on the behavior of beauty in the medium. The agreement with CMS data looks quite good.
\section{Conclusions and outlook}
We have presented predictions of our transport setup (both within a weak-coupling and a non-perturbative framework) for several heavy-flavour observables: non-photonic electrons, $D$-mesons and non-prompt $J/\psi$'s. Overall, concerning the quenching of the spectra, our results display a reasonable agreement with the experimental data up to momenta of order $\sim\!10$ GeV/c. On the other hand, even with the most extreme scenario for the thermalization time of the medium, it is hard to reproduce the observed amount of elliptic flow. Whether this has to be attributed to a too early decoupling of heavy flavour from the medium in our simulations (which could go on interacting in the hadronic phase), to a possible missing contribution received at hadronization (which in the presence of a medium might occur via coalescence) or to an incorrect modelling of the heavy-quark interaction with the plasma remains an open question.

As outlook of future work, a realistic modelling of in-medium heavy-flavour hadronization via coalescence with the light quarks from the plasma represents clearly a strong priority, both in order to assess its impact for the quenching and the azimuthal asymmetry of the spectra, but also in order to face the ongoing experimental efforts to study medium-modification of heavy-flavour hadrochemistry in $AA$ collisions. Our setup, tracking the evolution of each $Q\overline{Q}$ pair in the medium, will be also suited to the study of heavy-flavour correlations, currently at the center of a strong experimental interest, in particular in view of the perspectives opened by the upgrade program of the heavy-ion detectors. Finally the interface of our transport setup with full (3+1)D hydrodynamic codes will open the possibility of making predictions for heavy-flavour observables at forward rapidity.
\label{sec:conclusions}

\end{document}